\documentstyle[osa,manuscript,epsbox]{revtex}  % DON'T CHANGE %

\begin{document}                % INITIALIZE - DONT CHANGE % %  %

\title{Liquid-Solid Phase Transition and the Change of the Frictional Force of
the System with Two particles in a Box.}

\author{Akinori Awazu\footnote{E-mail: awa@zenon.ms.osakafu-u.ac.jp}\\ }

\address{Department of Mathematical Sciences\\
Osaka Prefecture University, Sakai 599-8531 Japan.         } %

\maketitle
\begin{abstract}
Liquid-solid phase transition and the
change of the frictional force of a system with two hard spheres in a
two-dimensional rectangular box are discussed. Under controlling the
pressure or the supply of energy from the wall, the solid like state,
the solid-liquid temporal coexistence state, and the liquid like state
are alternatively observed. The frictional force which 
works on particles and the mobility of the system are measured under the
supply of the energy with an asymmetric external force.
Characters of frictional forces for the $0$ mobility state and that for
the large
mobility state are obtained like, respectively, that of the static and the
dynamic frictions of solid-on-solid system.
The strong temperature dependency is also observed in the profile of the
relation between the above frictional force and the mobility.
From above results, the relation between the friction and the
velocity of a plate on granular layers which includes the hysteresis loop
[S.Nasuno et al.,Phys. Rev. Lett {\bf 79}  (1997) 949 et. al.] is
discussed. 
\vspace{2mm}
PACS number(s):\\

\end{abstract}

The appearance of the static friction and the dynamic friction, and the
change between them are familiar and important subjects of fundamental 
physics. They are universal phenomena which are usually observed at 
surfaces of macro-scale objects. 
From the microscopic point of view, the change between above two types of
frictions is expected to have close 
relationship with the liquid-solid phase transition around
surfaces\cite{sato00,sato0,nasuno,hayakawa,sato1}.
Here, the liquid-solid phase transition in this case is caused by variances of
external driving forces like shear forces, which means that this
transition occurs under non-equilibrium states.
By numerical simulations of systems
containing $10^{1\sim 4}$ hard or soft core particles, the liquid-solid phase
transition with a Van der Waals loop was
observed\cite{are21,are2,are211,are22,are31}. Very similar relations 
were observed also in the system with two hard spheres in a rectangular
box \cite{awa}. Moreover, this system showed not only the
solid-liquid phase transition but also the transition like glass
transition with the appearance of $\alpha$ and $\beta$ like relaxation
and the disappearance of the Van der Waals loop. 
From these facts, we can expect that the appearances of above phase
transitions are universal characters caused by the effect of
the excluded volume independent of the number of associating particles.
In this paper, we try to discuss the appearance of the static and the dynamic
frictions and the change between them through a simple model with a few
modifications of that introduced in the previous paper\cite{awa}. First, phase
transitions of the
system under equilibrium and non-equilibrium states by controlling the
supply of energy and the pressure in horizontal direction, are
investigated. This controlled pressure corresponds to the normal reaction
in the general treatment of friction. Based on these investigation,
characteristic features of frictional forces in this system are
discussed.

The system of our considerations consists of two-dimensional hard sphere
particles with unit mass and unit radius which are confined in a
two-dimensional rectangular box (Fig. 1). 
The right wall with unit mass of the box can move in the horizontal
direction. The position of the right wall is $X(t)$ $(X(t)>2)$ and the
constant force $-f$ is working on this wall in the horizontal direction. The
left wall is set at the origin of the horizontal axis and are connected to
the energy source. The bottom of the box is set at the zero height of the
vertical direction, and the position of the top of the box (the box
height) is $Y$ ($Y>4$). All walls are rigid, and interactions between
two particles, and, between a particle and a wall without energy sources, are 
only through hard core collisions. These tasks are implemented in the 
following manner; the tangential velocities to the collision plane are 
preserved, while the normal component of relative velocity $\Delta v_{n}$ 
changes into $-\Delta v_{n}$. A particle hitting the left wall connected to 
the energy source with the velocity $(v_{h},v_{v})$ bounces back with the 
velocity 
$(V_{h},V_{v})$ ($V_{h}>0$). Here, subscripts
$h$ and $v$ indicate, respectively, the horizontal and the vertical 
direction. The velocity 
$(V_{h},V_{v})$ is chosen ramdomly from the probability distributions 
$P_{h}(V_{h})$ and $P_{v}(V_{v})$ \cite{hh}
\begin{equation}
P_{h}(V_{h})=T^{-1} V_{h} exp (-\frac{V_{h}^{2}}{2 T})
\end{equation}

\begin{equation}
P_{v}(V_{v})=(2 \pi T)^{-\frac{1}{2}} exp (-\frac{(V_{v}-U)^{2}}{2 T})
\end{equation}
where $T$ is the temperature of the energy source. (We give the Boltzmann
constant as $1$.) The case with $U \ne 0$ means the asymmetric force
working on the system from the energy source in vertical direction. We
only consider the case of $Y>4$ which means these two
particles can exchange their positions in horizontal direction with each 
other. When $X(t)>4$, these spheres can exchange their positions in
vertical direction. On the contrary, these particles cannot exchange their
positions in vertical direction when $X(t)<4$. 
In the previous paper, we fixed $X(t)=X$ and defined the state with
$X>4$ as the liquid state and that with $X<4$ as the solid state\cite{awa}.
In this paper, however, because we will treat probability distributions of
$X(t)$ in the following discussions, we slightly modify definitions of
these states as follows; The solid state is defined as the state in 
which the probability distribution of $X(t)$ has a peak at $X(t)<4$, and
the liquid state is defined as the state in which the distribution has a
peak at $X(t)>4$.

First, we focus on the cases of $U=0$ which are the same as the
situation that a 
heat bath is connected to the left wall. Figure 2 is the probability 
distribution of the position of right wall $X(t)$ for, respectively,
when (a)$Y=4.5$, (b)$Y=5.0$, and (c)$Y=5.5$ for typical values $f$ under the 
same temperature $T$ ($=0.1$). 
Independent of $Y$, each distribution has only one peak at $X(t)<4$ for
large value of $f$ (solid state), or $X(t)>4$ for small value of $f$
(liquid state). 
For middle values
of $f$, however, each distribution has two peaks for the case of
$Y \le 5.0$; One of them is at $X(t)<4$ and the other is at $X(t)>4$
(Fig. 2 (a)). 
This indicates, in the present situation, the system passes between the liquid
state and the solid state. Thus, the discontinuous transition
between the solid state and the liquid state appears like in the system
with many hard spheres\cite{are211}. These two
peaks become blunt with the increase of $Y$, and the probability 
distribution around $X(t)=4.0$ becomes almost flat for $Y \sim 5.0$
(Fig. 2 (b)). Differently from the case of $Y \le 5.0$, each distribution
has always only one peak in the case of $Y>5.0$ (Fig. 2 (c)), where this
peak crosses the line $X=4$ with no singularities for the middle $f$.
The appearance and the disappearance of the discontinuous transition are
considered to be strongly related to those of Van der Waals loop which were
observed when the width of the box is fixed\cite{awa}.
However, the height with no discontinuous transition is $Y > Y^{*} \sim
5.0$, while the one with no Van der Waals loop is $Y' > Y'^{*} \sim 6.0$
in our previous paper\cite{awa}.
Our value $Y^{*}$ is, if anything, similar to another critical height
$Y'_{c} \sim 5.0$ for the previous model, where
$\beta$ and $\alpha$ like relaxations appear for $Y' \ge Y'_{c}$\cite{awa}.
If we fix the value $f$ and change the value
$\beta=1/T$, we observe qualitatively the same results as the above.   

Next, we discuss the cases with $U>0$. In these cases, the system becomes 
non-equilibrium because of the asymmetric force working on particles 
in vertical direction along the left wall. 
Figure 3 is the probability distribution of the position of the right wall
$X(t)$ for, respectively, when (a)$Y=4.5$ with low temperature ($T=0.1$),
(b)$Y=4.5$ with high temperature ($T=0.4$), (c)$Y=5.5$ with low
temperature ($T=0.1$), and (d) $Y=5.5$ with high temperature
($T=0.4$), for typical values $U$ under 
constant value $f$($=1$). Here, we set $T$ with which the probability 
distribution of $X(t)$ has only one peak for $X(t)<4$ (the solid state
is realized) for the case of $U=0$. 
The discontinuous phase transition between the solid state and the
liquid state with the liquid-solid temporal coexistence state is realized
for $Y \le 5.0$ by varying the value $U$ (Fig. 3 (a) and (b)). 
Differently from cases of the previous discussions with $U=0$, however,
such a discontinuous phase transition is observed also for cases with
$Y>5.0$. In these situations, two particles are compressed near the top
wall of the box by the asymmetric force with $U>0$. This means the region in
which particles spend almost all the time becomes smaller in vertical
direction. Then, the $Y$-dependent characters of the phase transition
become blunt. With increase of $T$ and $Y$, these two peaks become smoother
and the top of the probability distribution of $X(t)$ for middle $U$
comes close to be flat (Fig. 3(d)).

Now, we define the mobility $m$ of the system as the average frequency 
of the change of sign of $\delta y$. Here, $\delta y$ is the relative 
position between two particles in vertical direction. The solid state is 
realized when $m=0$, and the liquid state is realized 
when $m$ is large. From the asymmetric force
characterized by the velocity $U$, the force $F_{ex}=\lim_{t \to
\infty}\frac{1}{t}(\sum_{col}(V_{v}-v_{v}))$ works on particles at the 
left wall in average. Here, $\sum_{col}$ is the sum of individual 
collisions between the left wall and particles.
Figure 4 is the relation between (a)$U$ and $m$, and (b)$U$ and $F_{ex}$
for typical values $T$ under constant values $f=1$ and $Y=5.0$.
For small $U$, $m=0$ and $F_{ex} \propto U$ are realized, and 
$m \propto U$ and $F_{ex}= constant$ are realized for large
$U$. Here, $F_{ex}$ for each $U$ becomes large with $T$ decreasing, although
$T$ dependencies of $F_{ex}$ are small for large $U$.
Moreover, each $U$-$F_{ex}$ curve for small $T$ has a peak which
disappears for large $T$.
When the width of the box becomes larger, $m$ become larger.
Then, frequency of collisions between particles
and the left wall connected to energy source decreases. These facts make 
$F_{ex}$ to be almost constant for large $m$ although $U$ increases. 
Figure 5 indicates relations between $m$ and $F_{ex}$ by
increasing $U$ for typical values $T$ under constant values $f=1$,
where (a)$Y=5.0$ and (b)$Y=4.5$. Here, each state of the
system is almost stationary in which $m$ looks constant if averaged over
the macroscopic time scale sufficiently larger than the mean
free time of particles. 
Then, $F_{ex}$ is regarded as the driving force to keep each
steady state.
This also means that the
effective frictional force $R$, the magnitude of which is
equal to that of $F_{ex}$, emerges in the system to inhibit the
mobilization of the system. Hence, Fig. 5 also indicates the relations
between the mobility $m$ and the frictional force $R$.
Here, $f$ plays a role of the normal reaction of this system which is
treated as the constant $1$. Thus, the relations between
$m$ and $R$ are equivalent to those of $m$ and the effective friction
coefficient $\mu=\frac{R}{f}$. 
In Fig. 5, for small $T$, $R$ has a maximum value at a neighborhood of
$m=0$, and this decreases
monotonically and becomes almost constant with the increase of $m$.
These characteristics of $m$-$R$ relation are similar to those of 
velocity-friction relations of solid-on-solid system
which satisfies Coulomb-Amontons's frictions laws\cite{sato00}. 
For large $T$, on the contrary, $R$ increases monotonically and becomes
almost constant value with the increase of $m$ different from
Coulomb-Amontons's frictions laws. 
When $Y$ become small, $R$ increases a little with $m$ for large $m$
(Fig. 5 (b)). However, this increase is small, and qualitative
characteristics obtained in the above are considered to be independent
of $Y$. 

Figure 6 is the relations between, respectively, (a)$m$ and the friction
coefficient $\mu=\frac{R}{f}$, and (b)$f$ and $\mu$ by increasing
$f$ for typical values $T$ under a constant value $U$ ($Y=5.0$).
In this case, the liquid state with large $m$ and the solid state with
$m=0$ are realized for, respectively, small $f$ and large $f$.
In cases with $U=0$, large $f$ states are equivalent to small $T$ states. 
Hence, it is natural that profiles of the relations for small $m$ in
Fig. 6 (a) become similar to those of relations with small $T$ in Fig. 5.
Because of the same reason, the probability distribution of $X(t)$ for
large $f$ also behaves similarly to those with small $T$ in Fig. 3.
In Fig. 6 (a), $\mu$ of the $m=0$ states is obtained which is larger
than $\mu$ of the large $m$ state. Moreover, $\mu$ remains constant
independently of $m$ for large $m$.
In Fig. 6 (b), the derivative $\frac{d \mu}{d f}$ is small if $f$ is
small or large. This means
the frictional force is almost proportional to the normal reaction in
these region.  These characteristics are also similar to those of
solid-on-solid frictions which satisfy Coulomb-Amontons's frictions
laws\cite{sato00}. In Fig. 6 (b), for middle $f$, friction coefficients
for each $T$ increase with the increase of $m$. With the decrease of
$T$, the friction constant increases sharply. 

Now, we focus on the relation between the velocity of a plate on granular
layers and the friction between the plate and granular
layers\cite{nasuno,hayakawa}. Recently, from a sensitive experiment,
Nasuno et al. found that the relation between the plate velocity and the
frictional force makes a hysteresis loop\cite{nasuno}; The
frictional force is multi-valued, which is less for decreasing velocity
to $0$ than for increasing velocity from $0$. It is remarkable that a
loop, which looks very similar to above one, can be created by
combining $m$-$R$ relations for large and small $T$ (see Fig. 5). 
Thus, we try to understand qualitatively the mechanism of history
dependencies of velocity-friction relations on granular layers from the
$m$-$R$ relations in Fig. 5. Here, $m$ corresponds to plate velocity and
$R$ corresponds to friction on granular layers. We assume
that the granular temperature $T_{g}$ of the surface of granular layers
plays similar roles to the temperature of energy source in our
system. Here, $T_{g}$ is defined as the half of the mean-square of
velocity fluctuation of each granular particle. Initially, the plate and
each particle in granular layers do not move, where $T_{g}=0$. Soon after
the plate starts slipping, $T_{g}$ increases but it is expected to be
small at the early stage. Hence, the frictional force decreases with
increasing the plate velocity on granular layers like $m$-$R$ relations
with small $T$ in Fig. 5.
When plate moves faster, however, $T_{g}$ is expected to become large
because the plate excites each particle in granular layers. Then, the
frictional force is almost constant for finite plate velocity like $m$-$R$
relations with middle or large $T$ in Fig. 5. By this
friction, in the final stage, the plate stops and $T_{g}$ becomes $0$ again. 
Thus, in this granular system, $T_{g}$ varies with time and this
variance of  $T_{g}$ iterates all along because the plate is pushed
continuously. This iteration and the temperature dependency of the
frictional force explain the appearance of such a hysteresis loop in
granular materials.

In this paper, we discussed the liquid-solid phase transition and the 
occurrence of static and dynamic frictions of the system
with two particles in a rectangular box. First, we discussed the
liquid-solid phase transition by controlling the pressure in
horizontal direction or the temperature of a heat bath.
When the height of the box is small, the discontinuous liquid-solid phase
transition with the appearance of the liquid-solid temporal coexistence
was observed.
However, the liquid-solid phase transition becomes continuous when the 
height of the box is higher than a critical value. 
Next, we discussed the relation between the mobility and the frictional force 
of the system in which particles are excited by an asymmetric force in
the vertical direction. In the simulation, we observed the frictional
forces which are quite similar to the static friction and the dynamic
friction observed in solid-on-solid system. Moreover, we found that the
relation between these frictional forces and the mobility strongly
depends on the temperature of the heat bath. Taking these characteristics
into consideration,
we discussed the origin of the hysteresis loop in the granular 
friction. Our discovery of temperature dependency of these characters of
frictions is an important result of our simulation. More discussions
together with theoretical, numerical, and experimental studies are
needed in the future.
Also the analytical study of dynamical properties of frictional forces on
granular layers is left as an important issue. 

The author is grateful to H.Nishimori, K.Sekimoto, S.Nasuno, and
S.Yukawa for useful discussions. This research was supported in part by
Grant-in-Aid for JSPS Felows 10376.

\newpage

\begin{figure}[h]
\caption[]{Illustration of two particles system in a box with a moving
 wall (right) and energy sources (left)}
\end{figure}

\begin{figure}[h]
\caption[]{Probability distribution of the position of right wall
 $X(t)$ (PD) for typical values $f$ with, respectively, when (a)$Y=4.5$ (Middle
 $f=0.1725$), (b)$Y=5.0$ (Middle $f=0.1345$), and (c)$Y=5.5$ (Middle
 $f=0.1225$) under $T=0.1$ and $U=0$.}
\end{figure}

\begin{figure}[h]
\caption[]{Probability distribution of the position of right wall
 $X(t)$ (PD) for typical values $U$ and $f=1$ with, respectively, (a)$Y=4.5$ 
 with $T=0.1$ (Middle $U=1.14$), (b)$Y=4.5$ with  $T=0.4$ (Middle
 $U=0.665$), (c)$Y=5.5$ with $T=0.1$ (Middle $U=1.925$), and
 (d) $Y=5.5$ with $T=0.4$ (Middle $U=1.25$).}
\end{figure}

\begin{figure}[h]
\caption[]{Relations between (a)$U$ and $m$ and (b)$U$ and $F_{ex}$ for
 typical values $T$ ($T=0.06$, $0.24$ and $0.42$) with $Y=5.0$ and $f=1$}
\end{figure}

\begin{figure}[h]
\caption[]{Relations between $m$ and $R$ ($F_{ex}$) for
 typical values $T$ with, respectively, (a)$Y=5.0$ ($T=0.06$,
 $0.24$ and $0.42$) and (b)$Y=4.5$ ($T=0.05$, $0.21$ and $0.37$)
 ($f=1$)}.
\end{figure}

\begin{figure}[h]
\caption[]{Relations between (a) $m$ and $\mu$, and (b) $f$ and $\mu$,
 for typical values $T$ ($T=2^{-5}$, $2^{-4}$ and $2^{-3}$) with $Y=5.0$
 and constant $U$ ($U=2.0$).}
\end{figure}

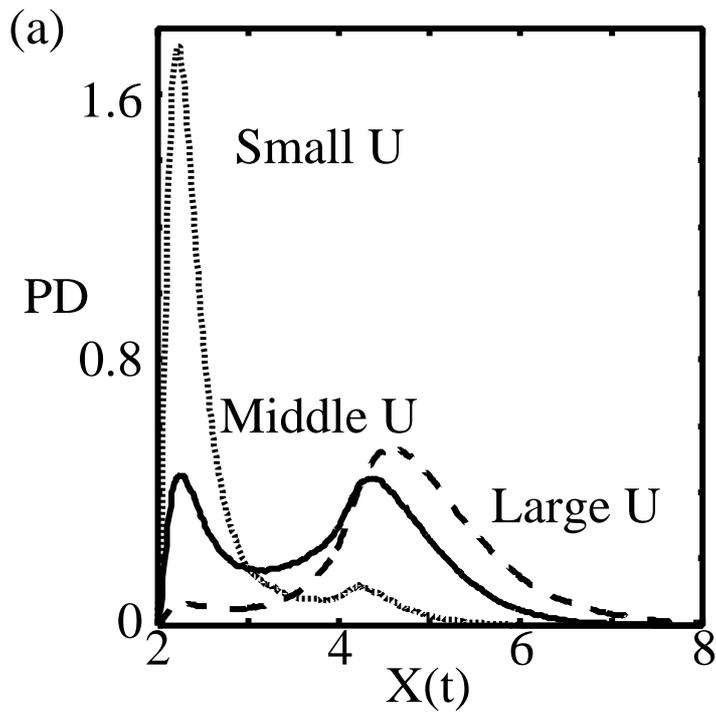
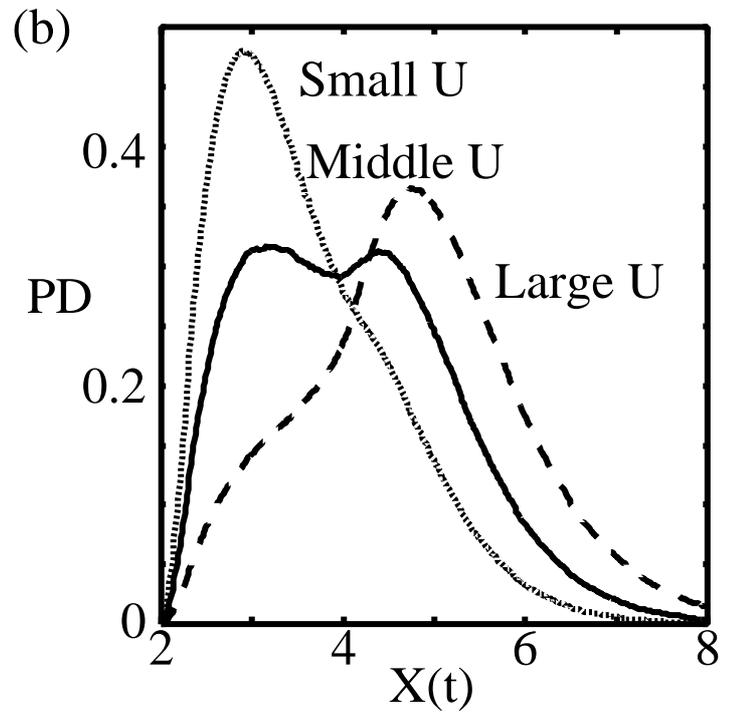
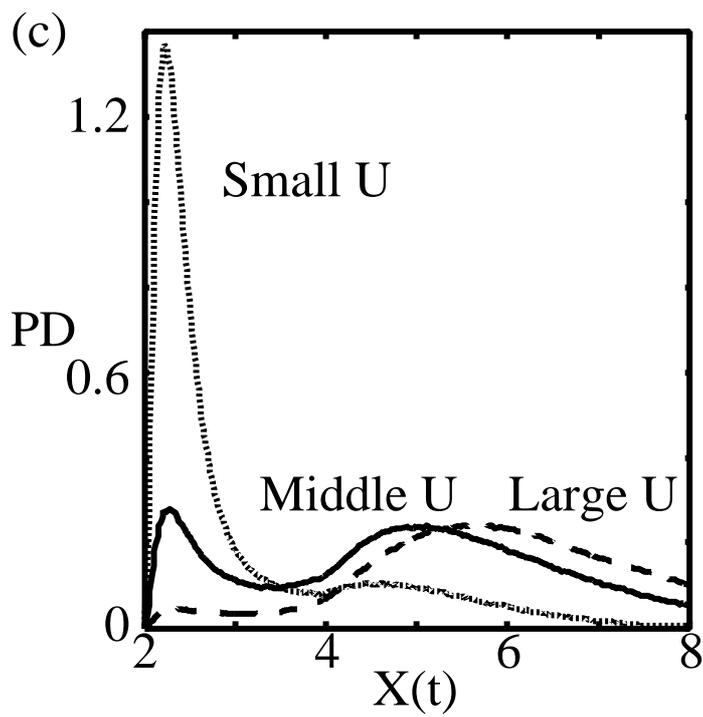
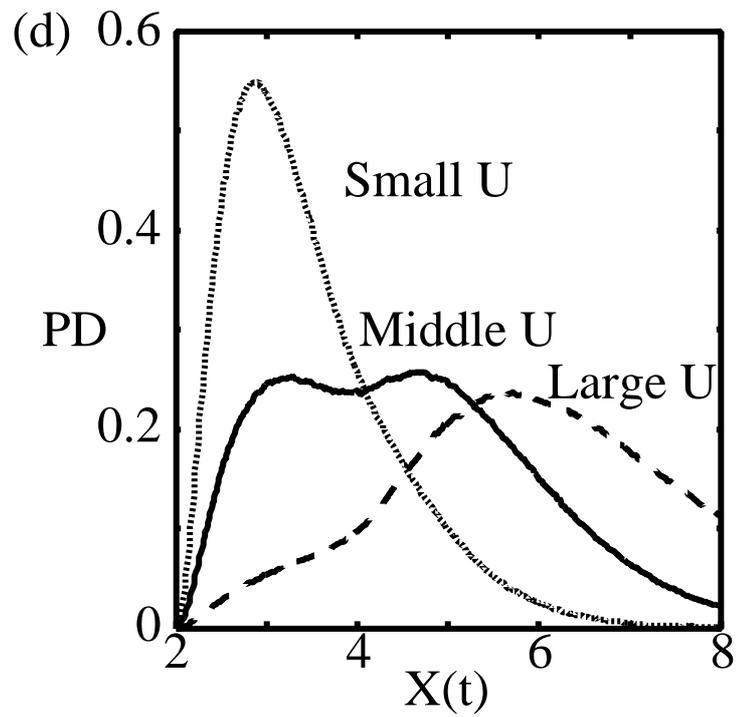
\end{document}